\documentclass[letter]{aa}  
\pdfoutput=1%
\usepackage{graphicx}
\usepackage{txfonts}
\begin{document}

\title{Observations of solar small-scale magnetic flux-sheet emergence}

   \author{C.E.~Fischer
          \inst{}
           \and
           J.M.~Borrero
           \inst{}       
             \and
           N.~Bello~Gonz\'{a}lez
           \inst{}      
                   \and
           A.J.~Kaithakkal
           \inst{}   
}

   \institute{Kiepenheuer Institut f\"{u}r Sonnenphysik, Sch\"{o}neckstrasse 6, 79104 Freiburg, Germany
}

   \date{Received ; accepted }

 
  \abstract
   {}  
   {\cite{2018ApJ...859L..26M} recently discovered two types of flux emergence in their numerical simulations: magnetic loops and magnetic sheet emergence. Whereas magnetic loop emergence has been documented well in the last years, by utilising high-resolution full Stokes data from ground-based telescopes as well as satellites, magnetic sheet emergence is still an understudied process. We report here on the first clear observational evidence of a magnetic sheet emergence and characterise its development.}
  {Full Stokes spectra from the Hinode spectropolarimeter were inverted with the SIR code to obtain solar atmospheric parameters such as temperature, line-of-sight velocities and full magnetic field vector information.}
   {We analyse a magnetic flux emergence event observed in the quiet-sun internetwork. After a large scale appearance of linear polarisation, a magnetic sheet with horizontal magnetic flux density of up to 194\,Mx/cm$^{2}$ hovers in the low photosphere spanning a region of 2 to 3 arcsec. The magnetic field azimuth obtained through Stokes inversions clearly shows an organised structure of transversal magnetic flux density emerging. The granule below the magnetic flux-sheet tears the structure apart leaving the emerged flux to form several magnetic loops at the edges of the granule. }
   {A large amount of flux with strong horizontal magnetic fields surfaces through the interplay of buried magnetic flux and convective motions. The magnetic flux emerges within 10 minutes and we find a longitudinal magnetic flux at the foot points of the order of $\sim$$10^{18}$ Mx. This is one to two orders of magnitude larger than what has been reported for small-scale magnetic loops. The convective flows feed the newly emerged flux into the pre-existing magnetic population on a granular scale. }
\keywords{Sun: photosphere -- Sun: magnetic fields}
\maketitle
%

\section{Introduction}\label{sec:intro}

Time series of magnetograms reveal patches of horizontal and vertical magnetic flux with apparent random distributions, which are then advected by the granular flows. {\cite{2018ApJ...859L..26M} have reported two distinct processes for magnetic flux emergence in their numerical simulations. In the magnetic flux tube emergence case, a magnetic loop emerges within a well established granule. The loop foot points, with predominantly vertical magnetic fields, are connected by horizontal magnetic patches. The foot points are then further separated and dragged to the intergranular lanes.  Observations of this case were shown for example in \cite{2007ApJ...666L.137C} and studied statistically in e.g. \cite{2017ApJS..229...17S}. \cite{2009ApJ...700.1391M} who analysed in detail the physical properties of small-scale loops found an emerged longitudinal magnetic flux in each foot point of on average 9.1\,$\times$\,10$^{16}$ Mx.  
The second scenario found by \cite{2018ApJ...859L..26M}  in their simulations is the emergence of a magnetic flux-sheet. In contrast to the magnetic loop emergence, sheet emergence is linked to the evolution and expansion (upward and sideways) of a granule with an organised horizontal magnetic flux-sheet spanning over an entire granule forming a mantle. These authors compared their simulations with the observations by \cite{2017ApJS..229....3C}. Those observations took place in a forming active region and involved several granules, leading to elongated abnormal granule structures. We report here of a magnetic sheet emergence event observed in the quiet-sun.

 \begin{figure*}
  \includegraphics[width=\textwidth]{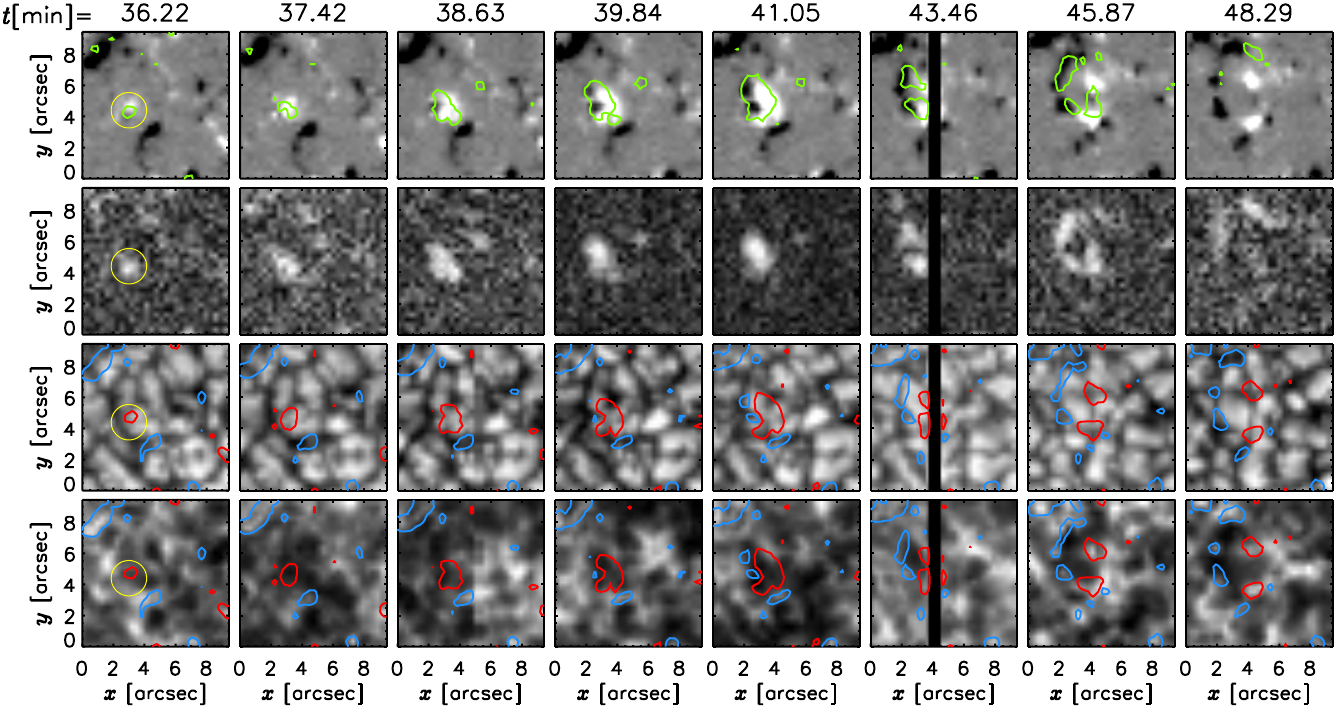}
  \caption{Temporal evolution of the emergence event. {\em Left to Right:} The top row shows the apparent longitudinal magnetic flux density with a cutoff at $\pm$ B$_{long}^{app}$=50 Mx/cm$^{2}$. The contours in green outline the apparent transverse magnetic flux density at a level of B$_{trans}^{app}$ of 140 Mx/cm$^{2}$. The second row shows B$_{trans}^{app}$ cut off at 320 Mx/cm$^{2}$. In the third row the continuum image is shown with contours of B$_{long}^{app}$ at a level of $\pm$ 30 Mx/cm$^{2}$ in blue and red respectively, whereas the last row shows the line core of the 630.15\,nm line with the same contours as in the third row. The yellow circle encompasses the initial flux emergence region. \label{time_ev}}
 \end{figure*}

 \begin{figure*}
   \sidecaption
  \includegraphics[width=\textwidth]{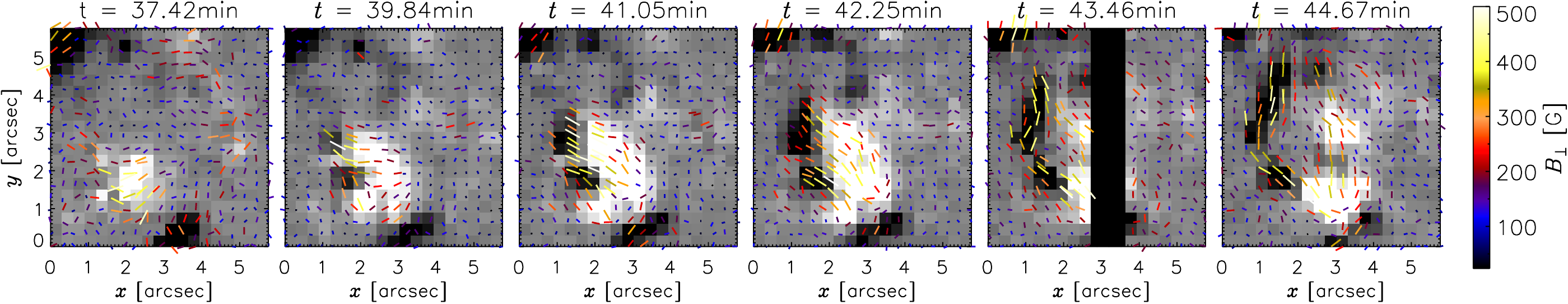}
  \caption{Horizontal magnetic field as derived from the SIR inversions. {\em Left to Right:} The background represents the apparent longitudinal magnetic flux density with a cutoff at $\pm$ B$_{long}^{app}$=50 Mx/cm$^{2}$ for a few of the time steps shown in Fig.~\ref{time_ev}, as well as additional frames. The lines indicate the magnetic field direction of the horizontal magnetic field obtained from the SIR inversions. Due to the $180^{\circ}$ ambiguity we do not show the arrow heads as they could be pointing in either of the two indicated directions. The colour of the lines indicate the strength of the total horizontal magnetic field. \label{az_ev}}
        \end{figure*}
 \begin{figure*}
   \sidecaption
  \includegraphics[width=\textwidth]{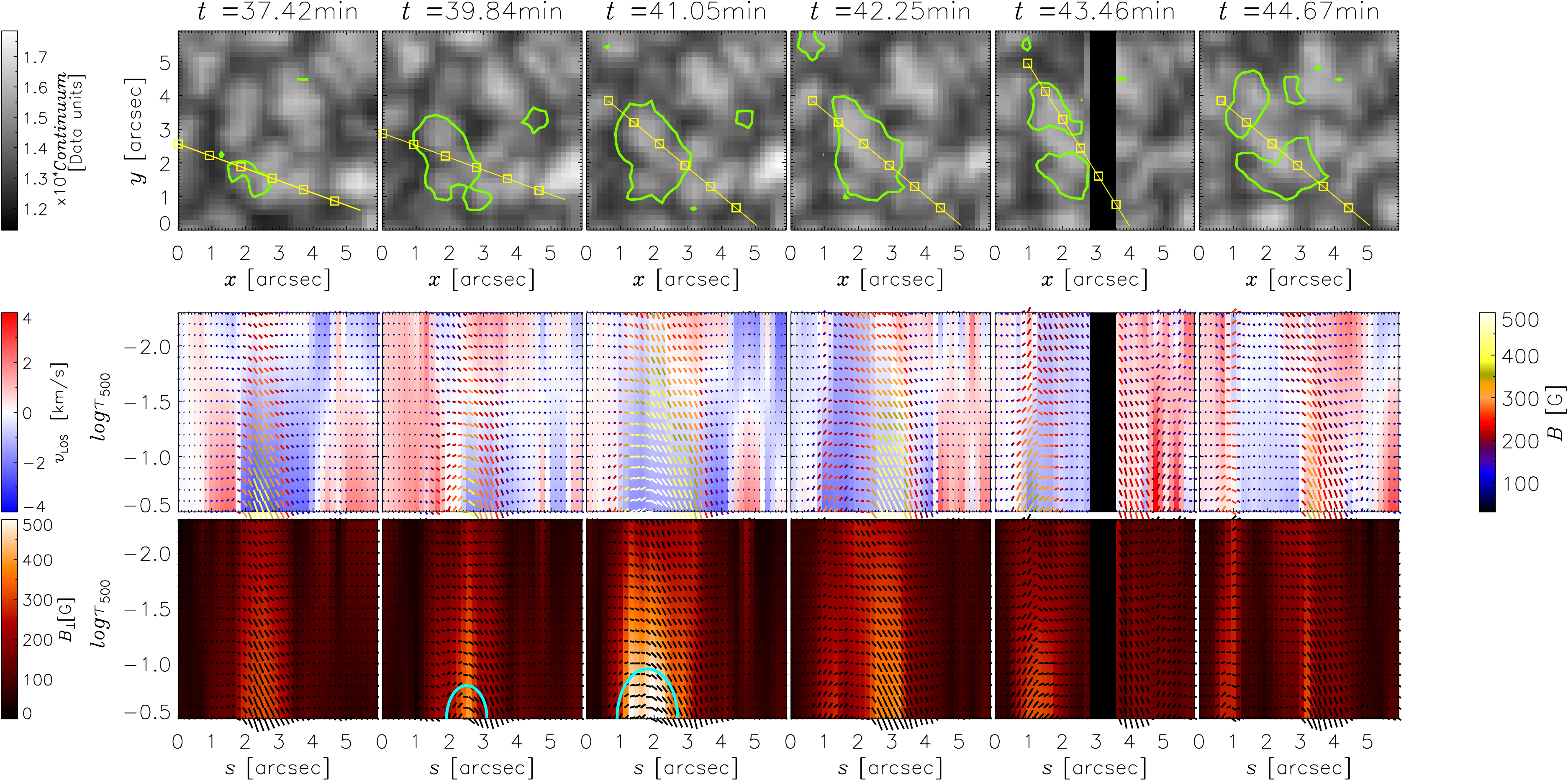}
  \caption{Continuum images, velocity, and magnetic field parameters with the time written above the first image of the column. {\em Top to Bottom:} The first row of images are 2-D maps taken in the continuum of the 630\,nm lines. The green contours correspond to the apparent transverse magnetic flux density with a cutoff at B$_{trans}^{app}$=140 Mx/cm$^{2}$. The yellow lines signify the cuts along a direction $s$ (running from left to right) with one arcsec intervals indicated by yellow boxes. In the second row the line-of-sight velocity is shown. Positive velocities correspond to down flows. On the $x$ axis, we follow along the cut $s$ shown in the first row and the $y$ axis displays the values at different heights measured in $\log\tau_{500}$. The coloured lines follow then the 2-D field map of the magnetic field vectors with the colours indicating the total magnetic field strength according to the colour bar on the right side. The third row shows the horizontal magnetic field strength obtained from the Stokes inversions. The lines following the magnetic field are now in black for better visibility. The cyan-coloured line, which outlines the loop shape, is to direct the eye. \label{crossy}}
        \end{figure*}

\begin{figure*}
\centering
   \includegraphics[width=17cm]{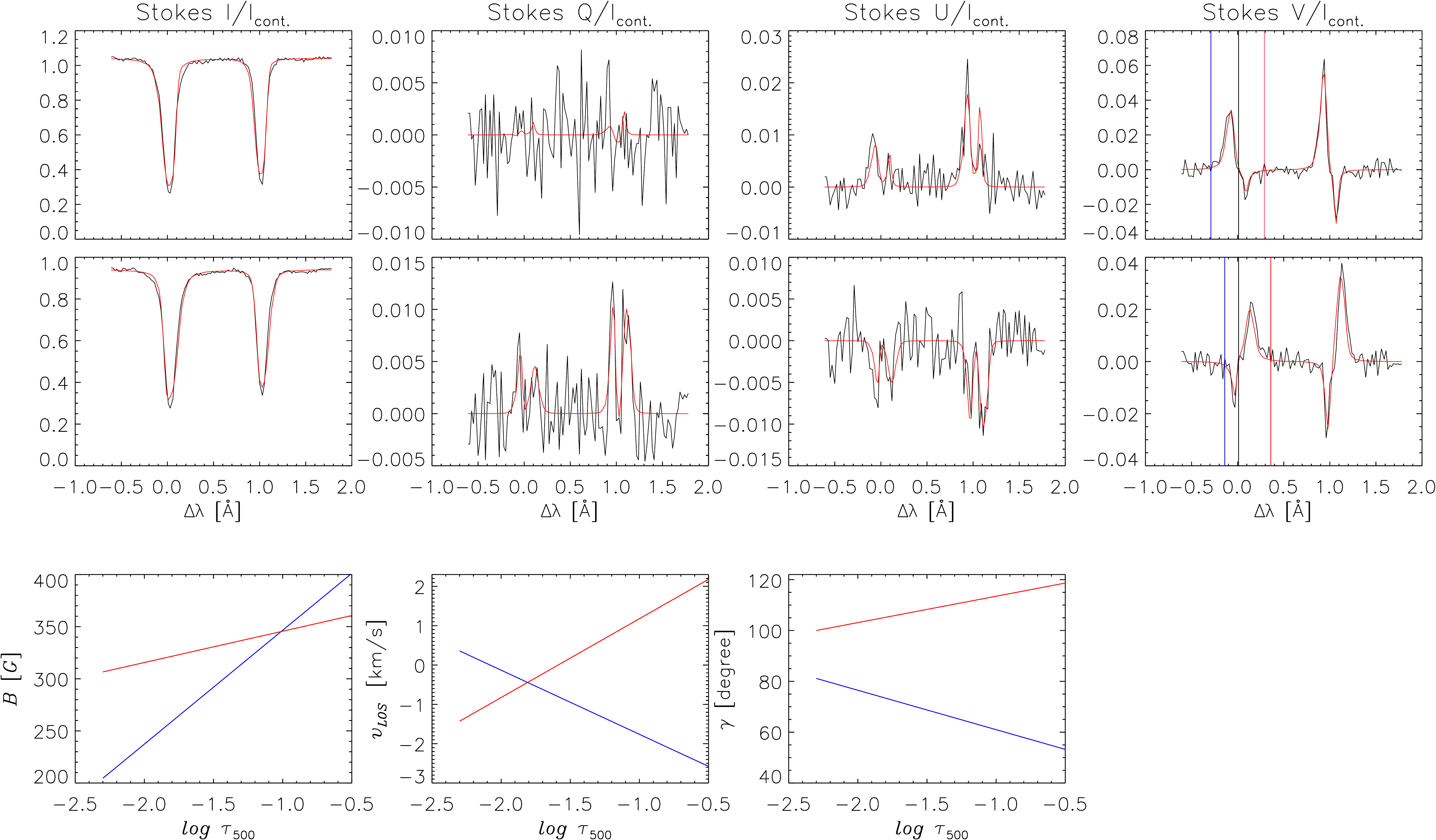}
     \caption{Stokes profiles and results from the SIR inversions. The upper row shows the results from the SIR inversions for a pixel located in an up flow region. The normalised observed Stokes profiles (black) as well as the fit (red) obtained from the inversions are shown. The wavelength scale corresponds to the $\Delta\lambda$ from the laboratory wavelength of the Fe\,{\sc{i}}\,6301.498$\AA$ line corrected for the gravitational redshift and convective blueshift (see Section~\ref{sec:obs}). The vertical black line plotted in the Stokes $V$ profile panel indicates the zero-crossing position, and the blue and red vertical  lines indicate the encompassed region  from which the area in the respective lobes are calculated.  The second row is in the same format as the upper row, but now for a pixel located in a down flow region.  The lower row shows the inversion results for the magnetic field strength ($B$), line-of-sight velocity ($v_{LOS}$), and inclination angle $\gamma$ for the blue-lobe dominant Stokes $V$ profile with blue solid lines and for the red-lobe dominant Stokes $V$ profile with red lines.  }
     \label{SIRPIX}
\end{figure*}


\section{Observations}\label{sec:obs}

The quiet-sun data set used in this study covers two and a half hours of near disc centre ($\mu=0.98$) observations and had been described in detail in \cite{2017A&A...602L..12F}.
Here we select only the $\emph Hinode$ Solar Optical Telescope~\citep{2008SoPh..249..167T}  data recorded by the Spectropolarimeter (SP) instrument. This slit scanning instrument records spectra in the 630\,nm wavelength band encompassing the Fe\,{\sc{i}}\,630.15\,nm and Fe\,{\sc{i}}\,630.25\,nm lines (effective Land\'{e}-factors of $1.67$ and $2.5$, respectively). The SP data was recorded in binned mode with a pixel size of about 0.32\,arcsec. The field-of-view resulted in about $9\times81$~arcsec$^2$, at a spectral sampling of 21.5\,m\AA/pixel, and a map cadence of $\sim$$\,70$\,s. We prepared level 1 data using the \emph{sp$\_$prep.pro} program \citep{2013SoPh..283..601L}. The output contains, amongst other parameters, the apparent longitudinal magnetic flux density and the apparent transverse magnetic flux density. 
Inversions were performed using the SIR code \citep{1992ApJ...398..375R} with 3 nodes in temperature, two nodes in magnetic field strength and inclination, as well as in line-of-sight velocity ($v_{LOS}$). Finally, only one node was employed in the azimuth of the magnetic field. This made it possible to account, for example, for gradients in the line-of-sight velocity and magnetic field. The inversions are performed using the laboratory wavelength of the spectral lines for the rest wavelength. We correct the retrieved velocity values by subtracting a correction value of 341\,m/s which accounts for the gravitational redshift and for the convective blueshift of -295\,m/s obtained by precise Lasercomb measurements by \cite{2018A&A...611A...4L}.

\section{Results}
\label{sec:res}

We follow the time evolution of the emergence event in Fig.~\ref{time_ev}. We found during the calibration process that data for a few scan positions were missing. This happened at ($t=43.46\,min$) and we mask the values in the images for those scan positions with zero. A movie of the event showing a smaller FOV, starting already at t=32.60\,min, and including overlays of the horizontal magnetic field can be found as online material ($event\_sheet\_emergence.mov$ see Fig.~\ref{snapmov}).\\
The first indication of the emergence is a strong apparent transverse magnetic flux density patch (B$_{trans}^{app}$) of 0.4\,arcsec$^{2}$ with the apparent magnetic flux density reaching up to  194\,Mx/cm$^{2}$. The horizontal field strength inferred from inversions is at a maximum of 296 G (at t=36.22\,min). At this point, positive polarity patches are observed towards the edge of the horizontal field. At t=39.84\,min, we observe negative polarity elements next to the positive polarity elements. By then, the horizontal field patch has also expanded in size. As more and more longitudinal magnetic flux emerges, the horizontal magnetic flux increases and develops into a large and coherent horizontal magnetic flux structure resembling a mantle while expanding sideways. At its maximum extent, the horizontal magnetic flux density region covers an area of $\approx$\,4.4\,arcsec$^{2}$, as large as the granule beneath it. At t=43.46\,min, the horizontal magnetic field structure seems to separate into two. Unfortunately, we could not retrieve the data at several scan positions. However, one can see clearly in the following time frame that the horizontal magnetic flux structure has now three distinct entities which move spatially farther apart later on. According to the continuum images (third row in Fig.~\ref{time_ev}), one finds a large granule expanding during this development. This granule seems to be responsible for the breaking up of the magnetic flux structure. The line-core images of Fe\,{\sc{i}}\,630.15\,nm seen in the last row of Fig.~\ref{time_ev} show the growing granule very clearly in the reversed granulation as an expanding dark blob surrounded by the longitudinal magnetic elements.\\
The lines showing the direction of the transverse field ($B_\perp$, given by the azimuth) in Fig.~\ref{az_ev} confirm the impression of a large and coherent magnetic flux mantle as the direction of $B_\perp$ remains the same over an extended area. One could argue that the breaking of the flux-mantle is only apparent and is due to its top part moving out of the line-forming region. However, a close inspection of the azimuthal direction of the magnetic field vectors as in, for example, t=44.67\,min, shows that positive and negative polarities are now connected at the sides of the former region of the magnetic flux-mantle. If the top of the flux-mantle had moved out of the line-forming region one would expect arrows pointing towards the middle (away from the middle) of the flux-mantle site from all directions. \\
We do not observe any abnormal granulation pattern. The granular flows are dominant and they push and separate the magnetic foot points and the mantle. \\ 
The emerged magnetic flux at the mantle foot points is estimated by tracking the emerging magnetic elements using a modification of the Multilevel-tracking-code (MLT) \citep{2001SoPh..201...13B} to account for negative as well as positive signals in the longitudinal magnetic flux maps. A threshold of 30\,Mx/cm$^{2}$, which corresponds to about 2.5 $\sigma$ of the signal in our maps, was chosen to track the magnetic elements. Once the emergence concluded the value of the total flux of the positive polarity foot points amounted to 1.4\,$\times$\,10$^{18}$ Mx. The emergence takes place within $9.6$\,minutes and we find therefore a flux emergence rate of 1.47\,$\times$\,10$^{17}$ Mx/min.\\
We obtained Stokes inversion results of which we show in Fig.~\ref{crossy} the line-of-sight velocities and magnetic field parameters in the range of $\log\tau_{500}$=-0.5 and $\log\tau_{500}$=-2.3, a good approximation of the formation height of the 630\,nm lines~\citep{2005A&A...439..687C}\footnote{$\log\tau_{500}$ refers here to the logarithm of optical depth for the $\lambda$=500\,nm wavelength}. 
We show cross sections chosen through the emerging feature at different times. For this, we consider a slice along the direction "s"  that connects the top-left and bottom-right opposite polarities of the flux-mantle. From the velocity values in the second row of Fig.~\ref{crossy} one can see that at t=37.42\,min, the horizontal flux emerges at a location of a granular up flow. The transverse field calculated as $B_\perp = B\sin\gamma$ from the inversion results (shown in the last row) reaches $500\,G$. The magnetic field direction is seen in the lines overlaid on the $v_{LOS}$ and transverse field images. The more vertical field lines identify the location of the apparent longitudinal magnetic flux density patch emerging almost co-spatially to the horizontal field patch.  At t=39.48\,min, a "loop-like" structure seems to form between $\approx\,s \in [1,3]$ arcsec. Cyan-coloured loops in the second and third frame of the last row in Fig.~\ref{crossy} should aid the reader's eye in identifying the "loop-like" structure. This can be interpreted as our cross section cutting through the horizontal magnetic flux-mantle. The positive polarity vertical field and the horizontal magnetic field are still located in the up flow region, whereas the negative polarity vertical field is at the edge of the granule and in the intergranular lane. At t=41.05\,min, the magnetic mantle has fully developed and one can clearly observe that the main horizontal magnetic field lies between $\approx\,s \in [1,3]$ arcsec encompassed by the vertical field. The horizontal structure covers the entire granular up flow region as seen from the velocity images, and it does not seem to rise further in height, staying below or at  $\log\tau_{500}$=-1.5 during the entire time sequence. At t=42.25\,min, the horizontal field strength in the last row has developed a gap with weaker field strength at around $s$=2 arcsec. The structure is being broken up and in the last time frame (last column) the horizontal magnetic field strength is confined to two locations at $s \approx 1$ and $s \approx 3$ arcsec. We do not observe anymore the loop structure, and the up flow region above the granule, which was covered before with horizontal magnetic flux, seems field-free.\\
We observe several pixels, within the emergence event time-sequence, where the circular polarisation (i.e. Stokes $V$) profiles show a blue lobe whose area can be very different from that of the red lobe. As those pixels exhibit also a net circular polarisation, the asymmetries must be caused by gradients in the atmospheric parameters along the line of sight. The animation provided as online material shows in the last panel locations of blue-lobe dominated Stokes V profiles in white and red-lobe dominated pixels in black. The entire emerging patch shows asymmetric Stokes $V$ profiles with larger blue lobes at the beginning of the time sequence. At the later time steps, the asymmetric profiles are located more on the edge of the structure, where the horizontal and longitudinal fields overlap. In general, more blue-lobe dominated pixels are located above the granules and red-lobe dominated at intergranular sites. This finding is consistent with the topology of a horizontal field region stacked above longitudinal foot points located at its edges.
 
In Fig.~\ref{SIRPIX}, representatives of the either blue-lobe dominated or red-lobe dominated profiles are shown. The pixel in the upper row of Fig.~\ref{SIRPIX} is taken during the first appearance of the emerging flux above a granule. We observe (as seen in the lower panel) an up flow in the deeper layers gradually decreasing and showing a slight down flow in the uppermost atmospheric layers at the edge of the formation layer of the Fe\,{\sc{i}}\,630\,nm lines. The magnetic field is around 400\,Gauss for the inclined magnetic field decreasing in strength with height and finally shows a weaker and more horizontal field at 200\,Gauss. For the pixel at the edge of the horizontal mantle, shown in the second row of Fig.~\ref{SIRPIX}, we have a similar situation in field strength, however, now the velocity gradient is reversed with down flows experienced in the lower atmosphere and gradually decreasing with height.

\section{Discussion}
\label{sec:disc}
Our observations show a similarity to the description of magnetic flux-sheet emergence in the quiet-sun simulations of \cite{2018ApJ...859L..26M}. In contrast to a small-scale magnetic loop emergence, the horizontal magnetic flux and the foot points of a magnetic sheet do not emerge within a granule in its late development, but develop with the granule. The horizontal magnetic flux-mantle we observe shows the topology of such a flux-sheet. We also observe an elongation of the longitudinal magnetic flux density patches at the edge of the flux-sheet during this process. This has been also seen in the simulation of magnetic flux-sheet emergence (private communication F. Moreno-Insertis). \cite{2018ApJ...859L..26M} found in their simulations a rate of 0.3 to 1 magnetic sheet emergence events per day and per Mm$^2$. The example we present here is the most prominent example of magnetic flux-sheet emergence we found in our (2.5 hours) quiet-sun time series. We can not make a statement yet on the importance of magnetic flux-sheet emergence for the magnetic flux emergence rate.\\
The dark blob we observe in the line-core images of Fe\,{\sc{i}}\,630.15\,nm is reminiscent of the magnetic bubbles of cool plasma described by \cite{2014ApJ...781..126O} in their observations of granular-sized magnetic flux emergence in an active region. They observe a dark bubble in the wing of the Ca\,{\sc{ii}}\,854.2\,nm line and follow its rise throughout the solar atmosphere. The bubble mimics the behaviour of the horizontal magnetic field patch during flux emergence and is located between the longitudinal foot points. In our observations, the dark blob seems to be rather a signature of reversed granulation, as it mimics the expanding granule and is also still visible once the horizontal magnetic flux-sheet has been torn apart.\\
\cite{2010A&A...511A..14G} studied a magnetic loop emergence in the quiet-sun and found an emerged magnetic flux of $\sim$\,3\,$\times$\,10$^{17}$\,Mx.
\cite{2007ApJ...666L.137C} reported a magnetic flux of the order of $\sim$\,10$^{17}$\,Mx for their example of magnetic flux emergence, and \cite{2009ApJ...700.1391M} found for the longitudinal magnetic flux an average of 9.1\,$\times$\,10$^{16}$ Mx at the foot points of their studied emerging magnetic loops. The event described in this letter carries 1.4\,$\times$\,10$^{18}$ Mx, therefore, a significantly larger amount of magnetic flux to the surface.
\section{Conclusion}\label{sec:conc}

We observe a case of emerging magnetic flux in the quiet-sun showing an organised transversal magnetic flux-sheet. The emerged magnetic flux is of the order of 10$^{18}$ Mx. The horizontal magnetic flux emerges together with the granule and the flux-sheet spans over the entire granular cell thereby "growing" with the expanding granule that carried the magnetic flux to the surface. To determine the contribution of magnetic flux-sheet emergence to the flux budget of the quiet-sun, statistics on these events are needed, which we will endeavour in the future. 
\begin{acknowledgements}
   CEF has been funded by the DFG project Fi- 2059/1-1 and CEF and AJK acknowledge financial support by the SAW-2018-KIS-2-QUEST project. JMB acknowledges financial support  by the Deutsche Forschungsgemeinschaft (DFG) - Projektnummer 321818926 (BO 4237/3-1). We thank Wolfgang Schmidt for valuable suggestions and comments. We thank the anonymous referee for their comments which helped to significantly improve the manuscript.
{\em Hinode} is a Japanese mission developed and launched by ISAS/JAXA, collaborating with NAOJ as a domestic partner, NASA and STFC (UK) as international partners. 
\end{acknowledgements}

\bibliographystyle{aa}   
\bibliography{em_lit}  

 \begin{figure*}
 \centering
\includegraphics[width=17cm]{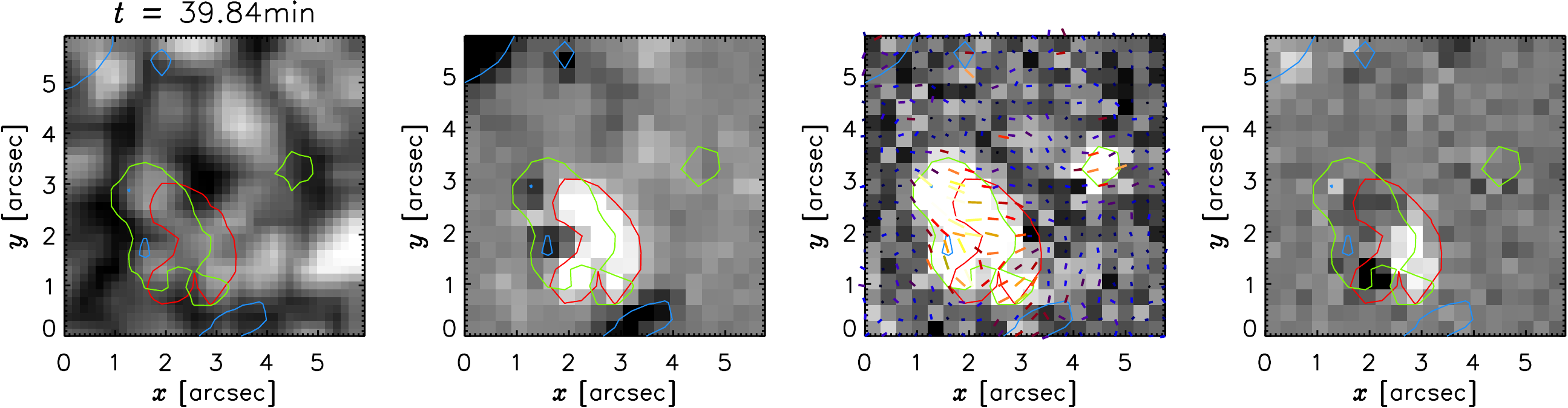}
  \caption{ Snapshot of the online movie $event\_flux\_emergence$.mov. {\em Left to Right:} The first image shows the continuum images with the time for the entire row indicated at the top of the first image. The second image is the B$_{long}^{app}$ at a cutoff level of $\pm$ 50 Mx/cm$^{2}$ followed by images in B$_{trans}^{app}$ at a cutoff level of 140 Mx/cm$^{2}$. The last image was obtained by subtracting the red lobe area in Stokes $V$ from the blue lobe area. All images show contours of B$_{long}^{app}$ at a level of $\pm$ 30 Mx/cm$^{2}$ in blue and red, respectively. The green contours correspond to the apparent transverse magnetic flux density with a cutoff at B$_{trans}^{app}$=140 Mx/cm$^{2}$. The lines overlaid on top of the third image starting from t=36.22\,min correspond to the horizontal magnetic field direction obtained from the SIR inversions. \label{snapmov}}
        \end{figure*}

\end{document}